# Defect Accumulation in $\beta$-Ga$_2$O$_3$ Implanted with Yb

Mahwish Sarwar[1], Renata Ratajczak[2], Cyprian Mieszczynski [2], Sylwia Gierałtowska[1], René Heller[3], Stefan Eisenwinder[3], Wojciech Woźniak[1], and Elżbieta Guziewicz[1]

[1] *Institute of Physics, Polish Academy of Sciences, Al. Lotnikow 32/46, 02-668 Warsaw, Poland*
[2] *National Centre for Nuclear Research, Soltana 7, 05–400 Otwock, Poland*
[3] *Helmholtz-Zentrum Dresden-Rossendorf, Institute of Ion Beam Physics and Materials Research, Bautzner Landstrasse 400, D-01328 Dresden, Germany*

Corresponding author:sarwar@ifpan.edu.pl

**Abstract**

Radiation-induced crystal lattice damage and its recovery in wide bandgap oxides, in particular beta-gallium oxide ($\beta$-Ga$_2$O$_3$), is a complex process. This paper presents the first study on the process of the defects accumulation in $\beta$-Ga$_2$O$_3$ implanted with Rare Earth (RE) ions and the impact of Rapid Thermal Annealing (RTA) on the defects formed. ($\bar{2}$01) oriented $\beta$-Ga$_2$O$_3$ single crystals were implanted with Yb ions fluences ranging from $1 \times 10^{12}$ to $5 \times 10^{15}$ at/cm$^2$. Channeling Rutherford Backscattering Spectrometry (RBS/c) was used to study the crystal lattice damage induced by ion implantation and the level of structure recovery after annealing. The quantitative and qualitative analyses of collected spectra were performed by computer simulations. The resulting accumulation curve reveals a two-step damage process. In the first stage, the damage of the $\beta$-Ga$_2$O$_3$ is inconspicuous, but begins to grow rapidly from the fluence of $1 \times 10^{13}$ at/cm$^2$, reaching the saturation at the random level for the Yb ion fluence of $1 \times 10^{14}$ at/cm$^2$. Further irradiation causes the damage peak to become bimodal, indicating that at least two new defect forms develop for the higher ion fluence. These two damage zones differently react to annealing, suggesting that they could origin from two phases, the amorphization phase and the new crystalline phase of Ga$_2$O$_3$.

## 1. Introduction

β-Ga$_2$O$_3$ is an extensively investigated wide bandgap (WBG) semiconductor with an energy bandgap of 4.5-4.9 eV [1]. The extremely high bandgap and chemical and thermal stability enable the high-temperature device operation for gallium oxide, outperforming renowned electronic materials in electronics such as GaN and SiC [2]. Monoclinic β-Ga$_2$O$_3$ is the most stable at room temperature and atmospheric pressure among other polymorphs including rhombohedral α-Ga$_2$O$_3$, spinel-like γ-Ga$_2$O$_3$, bixbyite δ-Ga$_2$O$_3$, orthorhombic ε-Ga$_2$O$_3$, and a transient form κ- Ga$_2$O$_3$ [3-7]. This compound has recently become a promising material for many applications and is gaining ongoing interest because of its potential use in various applications such as solar-blind photodetectors, UV transparent electrodes, field effect transistors, power electronics and others [8-10]. Specific potential applications for Ga$_2$O$_3$ are devices exposed to harsh environments, such as radiation-hard nuclear systems, space missions or high-energy applications. Exposure to all of these environments could create different types of crystal lattice damage in the crystalline material. Due to low symmetry and complex geometry of β-Ga$_2$O$_3$, the study of crystal lattice damage, creation, migration and transformation of defects upon irradiation is extremely complex [2].

Moving to the optical properties, β-Ga$_2$O$_3$, thanks to its high transparency, is a good host material for optically active centers in ultraviolet (UV) to infrared IR range of the spectral region. Undoped β-Ga$_2$O$_3$ emits light in the UV and visible (blue, green) region of the spectrum [11-13]. This range of luminescence can be further extended to the infrared by incorporating RE ions, particularly ytterbium (Yb). This aspect could open up tremendous possibilities for applications of β-Ga$_2$O$_3$ in various fields, especially in optoelectronics.

Ion-beam implantation is consistently used for the alteration of optical, electrical, and magnetic properties of semiconductors [14,15]. This technique works very well as an effective method for doping smaller areas and controlled doping of ions with low diffusivity coefficients, such as RE ions. However, it also has some disadvantages as due to the ballistic nature of this process, ion implantation causes lattice disorder leading to formation of defects and their transformation to increasingly complicated form of structural defects which might cause the shortening of the devices' lifetime and lead to the quenching of the luminescence in the irradiated host crystal or optical inactivation of the RE ion [16,17]. Fortunately, as shown for a range of materials, the crystal lattice can be recovered from the radiation-induced damage by using appropriate post-growth thermal treatment. However, for a new material such as β-Ga$_2$O$_3$, post-growth treatment must be optimized to achieve crystal recovery and optical activation of the RE ion. One of the effective methods for treating crystal lattice damage is Rapid Thermal Annealing (RTA) [18].

Although a few papers on post-implantation defects in β-Ga$_2$O$_3$ could be found in the literature, this issue still seems to be poorly understood, especially in case of gallium oxide bombarded with heavy ions. Azarov *et al.* [19] reported the accumulation of radiation disorder in (010) β-Ga$_2$O$_3$ single crystals implanted with Ni as a function of irradiation temperature and ion flux keeping the ion fluence constant at relatively low level no higher than 5 × 10$^{12}$ at/cm$^2$. The results obtained by channeling Rutherford Backscattering Spectrometry (RBS/c) technique showed lower amorphization in the crystals with an increase of irradiation temperature from 25 to 300°C and a lower level of disorder for the lower flux. Kjeldby *et al.* [20] studied the orientation, ion fluence, and annealing temperature dependent defect accumulation in Si-implanted β-Ga$_2$O$_3$. The research pointed at significant strain accumulation in the implanted crystals which results in crystal-to-crystal transition instead of amorphization and complex defect transformations after annealing. Lopez *et al.* [21] studied the correlation between crystalline quality and the luminescence of RE implanted β-Ga$_2$O$_3$ nanowires by implanting RE ions with a high ion fluence of $5 \times 10^{15}$ at/cm$^2$. After high temperature annealing, a significant recovery of the crystal lattice was inferred from Raman spectroscopy and Transmission Electron Microscopy (TEM) but, due to nanowire structure of the studied object, any RBS/c data could not be provided. Lorenz *et al.* [22] reported the defect formation and post-annealing recovery in Eu implanted β-Ga$_2$O$_3$ bulk crystals and nanowires with (201) orientation for a few fluences between $1 \times 10^{13}$ and $4 \times 10^{15}$ at/cm$^2$. They found that surface amorphization starts at a fluence of $4 \times 10^{15}$ at/cm$^2$ and proceeds deeper into the crystal lattice for higher fluence. They also reported that defect recovery starts at annealing temperature of 700°C, but it is more effective at even higher temperature.

Detailed studies of the defect accumulation process itself in (010) β-Ga$_2$O$_3$ implanted with P, Ar and Sn ions have been reported by Wendler *et al.* [23]. This paper presents two or three steps of the accumulation curves, but the amorphization level has not been achieved. However, it should be emphasized that the defect accumulation process in β-Ga$_2$O$_3$, as well as the created defect types, strongly depend on the crystal orientation as well as the physical and chemical properties of the dopant ions implanted in the crystal lattice [10,22,24].

This brief review shows that the radiation-induced defect accumulation process in single crystalline β-Ga$_2$O$_3$ was partially studied for lighter ion implantation only, while for gallium oxide implanted with rare earth ions neither the defect accumulation process nor the influence of post-annealing on the formed defects and the structure recovery have been elaborated, although this knowledge is necessary for future optoelectronic applications. The aim of the present study is to fill this gap.

In this paper, we investigate Yb-implanted $(\bar{2}01)$ oriented β-$Ga_2O_3$ single crystals by applying eleven ytterbium fluences ranging from $1 \times 10^{12}$ to $5 \times 10^{15}$ at/cm$^2$. Such a high number of fluences allows for a detailed study of the post-implantation defect accumulation process. Experimental RBS/c is used to determine the structural changes in the Yb-implanted and annealed $(\bar{2}01)$ oriented β-$Ga_2O_3$ crystals. Experimental results are supported by a range of advanced computer simulation programs such as SRIM, SIMNRA, and McChasy, used for the quantitative analyses of the defect accumulation process.

**2a. Experimental conditions**

A commercially available wafer of undoped single crystalline β-$Ga_2O_3$ (0.68 mm thick) with $(\bar{2}01)$ surface orientation from Tamura Corporation was used for the studies. Ion implantation of Yb with fluences ranging from $1 \times 10^{12}$ to $5 \times 10^{15}$ at/cm$^2$ and energy 150 keV, was accomplished at Helmholtz-Zentrum Dresden-Rossendorf, Germany (HZDR). RTA was performed at 800°C, in an oxygen atmosphere for 10 min by using an Accu Thermo AW-610 from Allwin21 Corporation system at the Institute of Physics, Polish Academy of Sciences (IP PAS). The standard RBS/c measurements were performed, with 1.7 MeV He$^+$ ions using a Van de Graaff accelerator at HZDR. In the RBS/c experiments, the silicon detector positioned at a scattering angle of 170° was used, with a depth resolution < 5 nm and an energy resolution < 20 keV. The surface morphology was investigated by Atomic Force Microscopy (AFM), Bruker Dimension Icon, using silicon nitride probes with sharp tips (a tip radius: 2 nm) in the Peak

Force Tapping mode at the IP PAS. The surface roughness was evaluated as root mean square (RMS) of the AFM height measurements. The RMS value for all samples was detected for scanning areas of 10 x 10 μm.

### 2b. Theoretical calculations details

SRIM calculation was used for theoretical estimation of the ion range and defect depth distribution for the defects. The collected RBS/c spectra were analyzed by computer simulation programs such as SIMNRA and McChasy supported by the SRIM program developed by Ziegler and Biersack [25]. The latter is a theoretical program based on the Monte Carlo Simulation method that calculates depth distributions of simple defects (displaced atoms and vacancies), as well as generates information about the stopping power and range of ions bombarded into matter. It is a reference program for the preliminary estimation of radiation effects, based on binary collision approximation with random selection of the impact parameter of the next colliding ion. The calculations provide information on stopping power, range of ions, and straggling distributions for any ion at any energy (in the range of 10 eV- 2 GeV) and for any elemental target. SIMNRA [26] is a simulation program for developing spectra of Nuclear Reaction Analysis, RBS, and Elastic Recoil Detection Analysis in the random mode. It was used to find out the RE concentration and its depth profile from the RBS random spectra. Quantification of defect concentration and its depth profile based on the obtained RBS aligned spectra (in the channeling mode) is typically done by Two-Beam Approximation [27]. However, the results obtained by this method could be misleading due to the fact that ion implantation into semiconductor compounds usually results in the creation of a mixture of defects [28-30]. Therefore, in order to estimate depth profiles and concentrations of defects, in the quantitive and qualitative analyses of collected RBS/c spectra, the Monte Carlo simulation code called

McChasy was used due to its unique ability to distinguish contributions to the aligned spectra coming from both simple and extended defects. In McChasy, a simulated RBS spectrum is generated by recreating the movement of ions in the solid. According to nuclear encounter probability, the scattering probability is calculated for each atom-ion interaction and is used in analysis of the RBS spectrum. The depth distribution for different types of defects: simple defects (RDA) such as randomly displaced atoms (or others resulting in direct scattering) and extended defects (DIS) such as dislocations (or others resulting in bending of channels) are deduced from this code. This program makes it possible to follow the evolution of different defects in the material as well [30].

The initially predicted depth distributions of ions and defects calculated by SRIM for 50000 incident particles of Yb with energy 150 keV incident on $\beta$-$Ga_2O_3$, show the calculated ions range of ~30 nm with the maximum of the nuclear energy loss profile ($R_{pd}$) at ~20 nm, while the whole thickness of the modified layer reaches up to 80 nm (Figure 1). SRIM calculations of vacancies caused by Yb ions were also used to determine the number of displacements per Yb atom (dpa). This number describes the fluence in a more general way, according to the approach presented elsewhere [31]. In all calculations, the density of β-Ga$_2$O$_3$ was set as 5.88 g/cm$^2$, and displacement energies of Ga and O atoms as 25 and 28 eV, respectively [32].

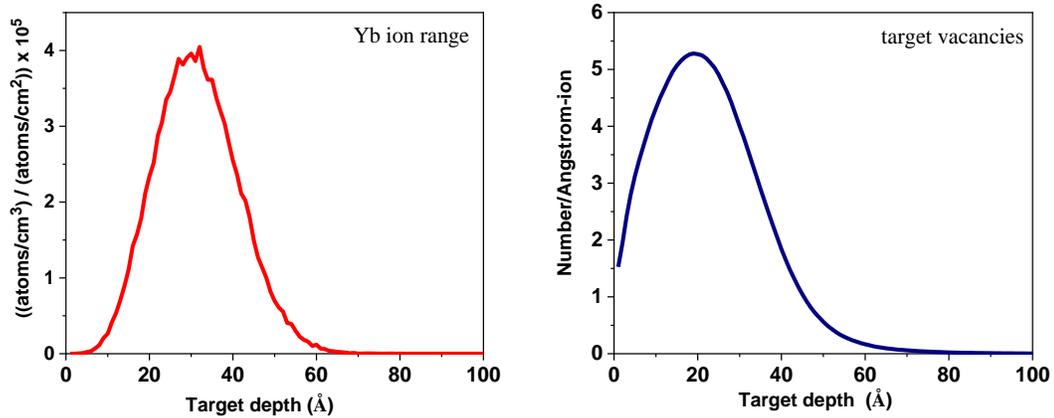

**Figure 1.** Ion range and the defect depth profile calculated by SRIM

## 3. Results

The virgin β-Ga$_2$O$_3$ crystals used for the implantation were atomically flat with surface roughness of 1.3 nm indicated by RMS (Figure 2(a)). After ion implantation with Yb fluence of 1 × 10$^{14}$ at/cm$^2$, the roughness decreased to 0.7 nm (Figure 2(b)) and eventually decreased to 0.6 nm after the performed annealing (Figure 2(c)). It is reported that decrease in surface roughness i.e. smoothening of the surface after a particular ion fluence may indicate amorphization of the surface [33].

It is worth noticing that this process is opposite to that observed in ZnO, where increase of RMS is reported after each implantation fluence. For example, Ratajczak *et al.* [18], reported a slight increase of roughness in the Yb implanted ZnO but the ion fluence used was different (5 × 10$^{14}$ at/cm$^2$) in that case. After the annealing, decrease in the roughness was observed which was also demonstrated in our case. Surface smoothness after annealing could be explained by surface stress relaxation [33].

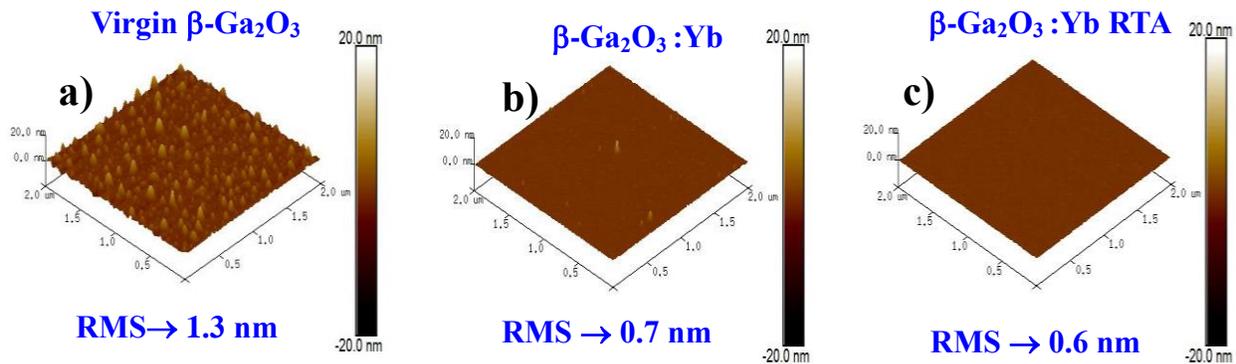

**Figure 2.** AFM images of a) virgin β-Ga$_2$O$_3$, b) β-Ga$_2$O$_3$ implanted with Yb fluence of 1 × 10$^{14}$ at./cm$^2$ and c) β-Ga$_2$O$_3$ implanted and annealed in oxygen at 800°C for 10 min.

**Defect accumulation after implantation**

The random and ($\bar{2}$01) aligned RBS spectra for β-Ga$_2$O$_3$ implanted with a medium energy of 150 keV and different Yb-ion fluences exhibit the Ga signal, obtained from He ions backscattered from Ga atoms, in the lower energy part of the spectra (1100-1400 keV) (Figure 3). The evolution of the aligned RBS spectra in this region reflects the increase in radiation-induced damage to the material's crystal lattice. For lower fluences, from $1 \times 10^{12}$ to $5 \times 10^{12}$ at/cm$^2$, the intensity of the aligned spectra is similar to the virgin sample. Starting from the Yb ion fluence of $1 \times 10^{13}$ at/cm$^2$, the radiation-induced damage peak observed between 1310 and 1335 keV, begins to increase rapidly that is accompanied by an increase of the dechanneling level, and for a fluence of $1 \times 10^{14}$ at/cm$^2$, the damage reaches a random level. But for the next used fluence ($4 \times 10^{14}$ at/cm$^2$), the damage region starts to spread with increasing ion fluence and the damage peak becomes clearly bimodal (showing two maxima), and their level is now lower than random. Further irradiation led to the saturation of the damage peaks at the random level again and the intensive evolution of the dechanneling level. These observations indicate the complexity of the defect accumulation process in the RE-irradiated β-Ga$_2$O$_3$ crystal (see Figure 1 in SI). The appearance of a bimodal peak is consistent with the literature and has been reported by Lorenz *et al.* [22].

In the higher energy part of the RBS spectra (1450-1600 keV), the signals from the Yb dopant appear. For fluences below $1 \times 10^{14}$ at/cm$^2$, the Yb signal is hardly visible. For the higher-used fluences (above $1 \times 10^{14}$ at/cm$^2$, the RE signal starts to grow and from the ratio of random and aligned spectra, it can be stated that for these fluences and along the ($\bar{2}$01) direction, the RE ions occupy the interstitial positions in the β-Ga$_2$O$_3$ crystal lattice.

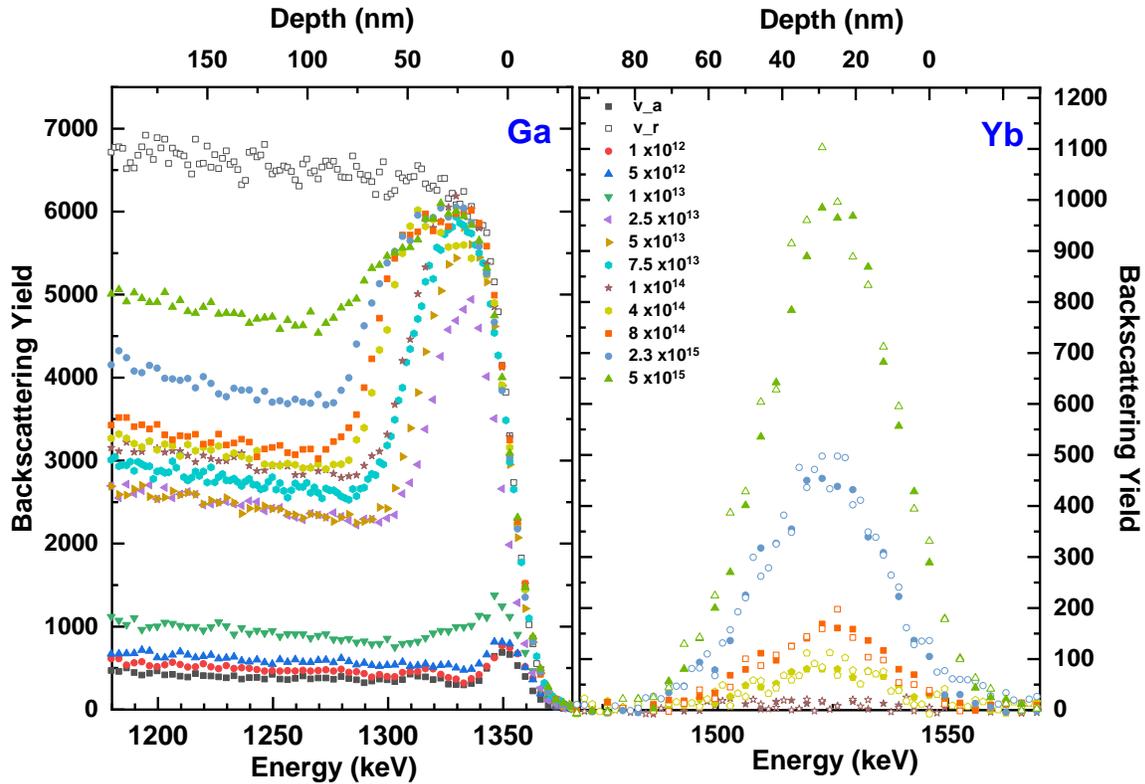

**Figure 3.** Random (open symbols) and aligned (solid symbols) RBS spectra for $(\bar{2}01)$ β-$Ga_2O_3$ prior to and after implantation with different fluences of the Yb ions.

The McChasy simulations were involved for the detailed structural defects analysis represented by solid lines (Figure 4). In order to better visualize the evolution of the damage peak, the shifted RBS/c spectra are also shown in Figure S1 (Supplementary Material, SM). The best fit between simulated and experimental spectra was obtained for the depth distribution of RDA and DIS type of defects shown in Figure 5. The defect accumulation curves based on simulated damage profiles are plotted in Figure 6. Points of the curves in Figure 6 correspond to the total number of RDA-type of defects at the 0-20 nm depth region from RDA distributions and of DIS-type of defects from the 40-60 nm of DIS distributions (Figure 5). The upper x-axis in Figure 6 reflects the ion fluence calculated as dpa. Solid lines in Figure 6 are the fits made following the MSDA model [34].

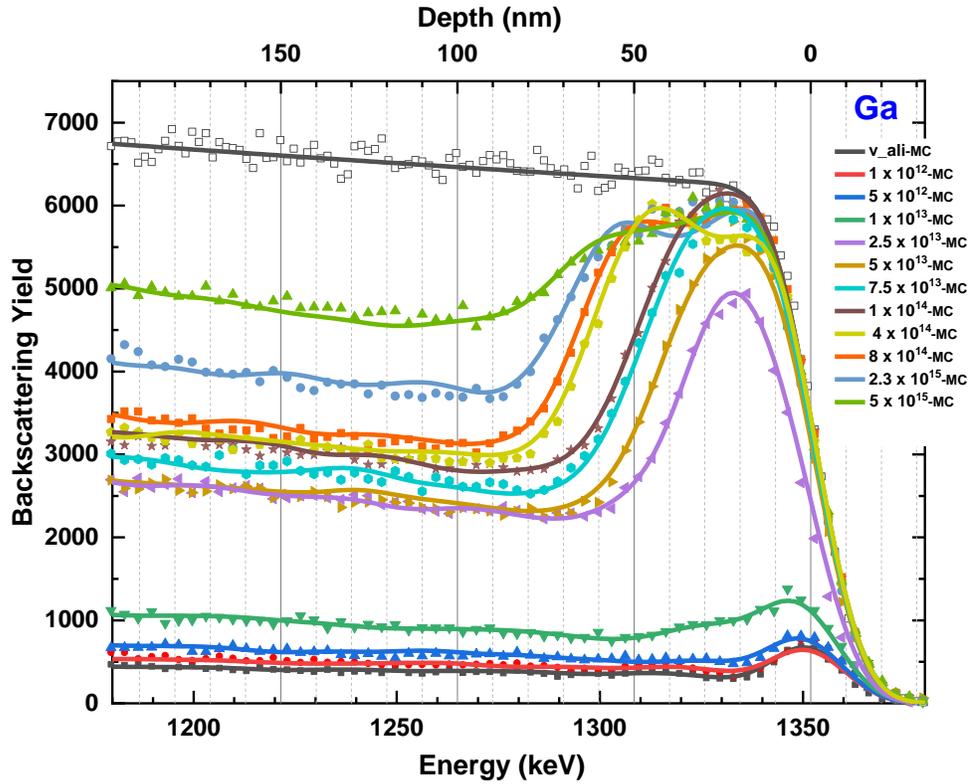

**Figure 4**. Experimental random (solid symbols) and aligned (open symbols) RBS spectra for β-Ga$_2$O$_3$ prior to and after implantation with different fluences of the Yb ions. The solid line represents the results of the McChasy simulation.

Based on Figure 6, four regions of damage buildup can be distinguished in Yb-bombarded β-Ga$_2$O$_3$. For the lower Yb fluences (up to $5 \times 10^{12}$ at/cm$^2$, i.e. region I), the concentration of both types of defects is relatively low. This effect has been usually attributed to dynamic annealing, i.e. defect migration, recombination and clustering during irradiation [19]. Starting from the Yb fluence of $1 \times 10^{13}$ at/cm$^2$, the damages increase very fast, reaching the saturation level of approximately $23 \times 10^{10}$ of DIS type of defects /cm$^2$, and 100% of RDA type of defects for the Yb fluences of $1 \times 10^{14}$ at/cm$^2$ (region II). Such rapid changes in the accumulation curves are usually associated with defect structure transformations [23,27-29].

In region III (from $1 \times 10^{14}$ to $8 \times 10^{14}$ at/cm$^2$ of Yb fluence), the concentrations of both types of defects are constant. However, for the fluence equal to $4 \times 10^{14}$ at/cm$^2$, one unexpected and very important change in the shape of the damage peak can be noticed (Figure 4), reflected in the modeled RDA defect profile (Figure 5). Namely, the damage peak visible in the spectrum became bimodal with a clear visible gap between two, well distinguished RDA-type damage zones, with the intensities of the two constituents not equal. What is important, the concentration of RDA in the damage zone closer to the surface is less than 100% now. It should be also noticed that this gap, visible also in the spectrum (Figure 3), appears in the depth region corresponding to the Yb 150 keV ion range (30-40 nm). However, it cannot be assigned to the layer stoichometry changes, or even to the presence of Yb itself because this issue is already taken into account in the McChasy code. Additionally, if this had been the case, this mechanism should have led to a deeper gap for higher irradiation fluence, but such a phenomenon is not observed in the subsequent spectra. The further irradiation leads to the saturation of RDA-type defects in the damage zone nearest to the surface at 100% again, and its broadening, while the deeper RDA damage zone seems to disappear. Above the Yb fluence of $8 \times 10^{14}$ at/cm$^2$ (region IV), the concentration of DIS increases very fast again, indicating the changes in the defect structure of extended defects in Ga$_2$O$_3$ irradiated with high fluences of RE.

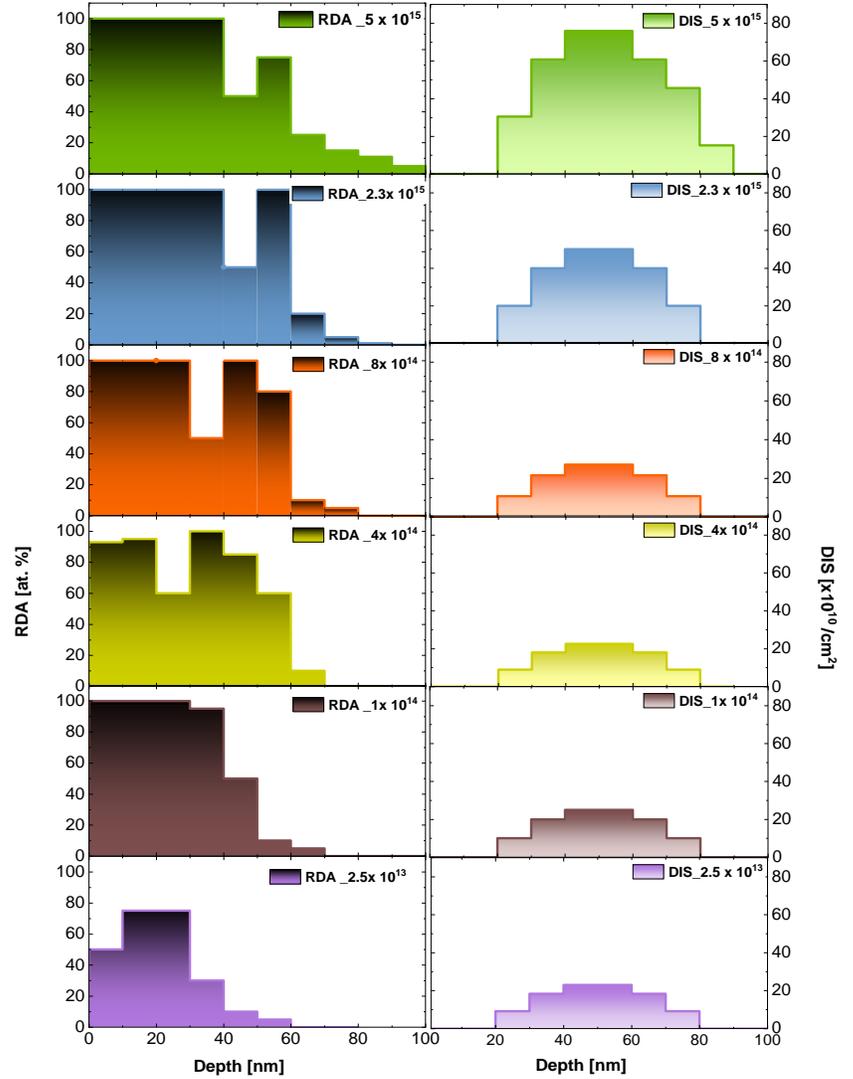

**Figure. 5.** Distribution of Randomly Displaced Atoms (RDA) and Dislocations, extended defects (DIS) defects for Yb implanted β-Ga$_2$O$_3$ obtained by McChasy

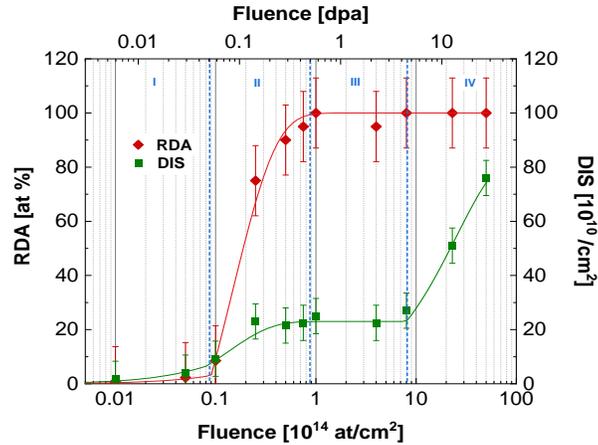

**Figure 6.** The accumulation curves of simple (RDA) and extended (DIS) types defects obtained by McChasy simulations. For the points, the standard error values were added, and the MSDA fits are plotted.

**Lattice Structure Recovery**

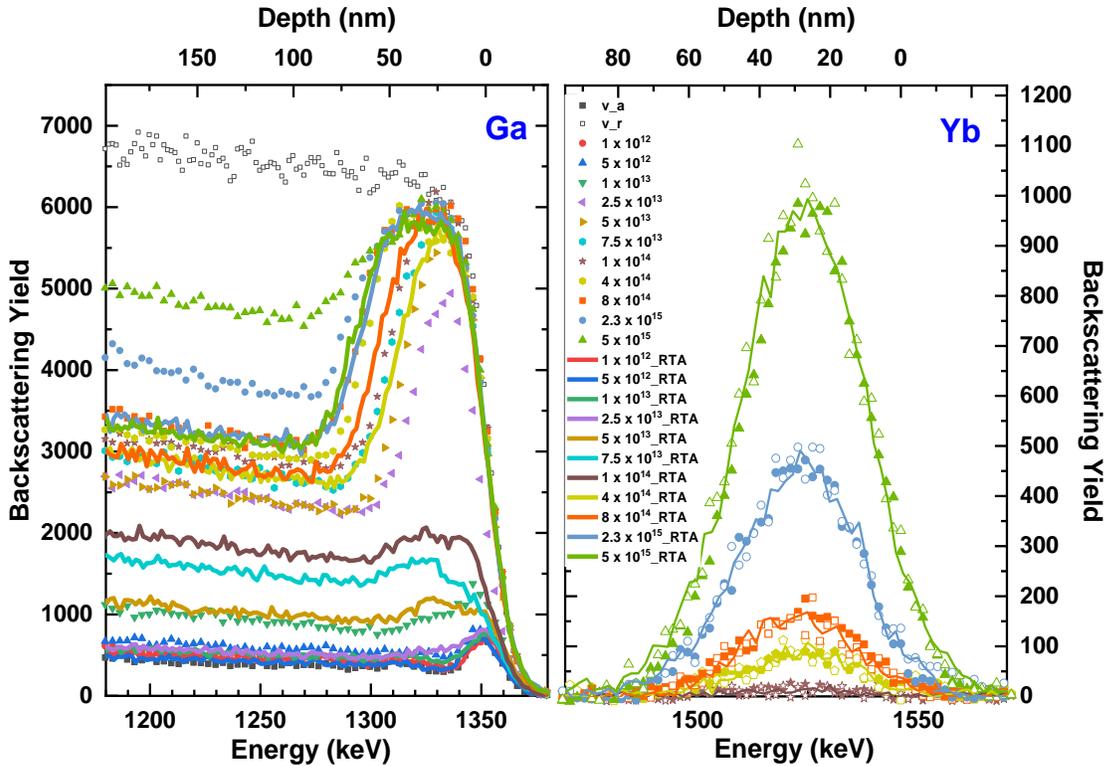

**Figure 7.** Random (ran) and aligned (ali) RBS spectra for β-Ga$_2$O$_3$ implanted with Yb and post-annealed at 800 °C, in O$_2$ for 10 min

The random and ($\bar{2}$01) aligned RBS spectra for β-Ga$_2$O$_3$ implanted and post-annealed at 800°C, in O$_2$ for 10 minutes was studied for elaboration of the crystal lattice recovery after thermal treatment (Figure 7). Such annealing conditions using the RTA system have been found to be the most effective for the crystal lattice recovery and optical response of RE ion in the ZnO crystal lattice [18]. After annealing, the recovery of the crystal is observed as a reduction in the damage peak for fluences not higher than $1 \times 10^{14}$ at/cm$^2$ (see also Figure 8). The damage caused by the higher fluences appears to be resistant to annealing, which can be attributed to the plastic deformation of the crystal. However, as can be seen by looking at the Ga signal in Figure 7, the thickness of the damage zone created by the higher fluences, as well as the dechanneling level, become lower after thermal treatment.

Looking at the Yb signal in Figure 7, it can be noticed that the distributions of Yb after annealing remain unchanged, i.e. Yb atoms stay at interstitial positions, and the out-diffusion of Yb-ions after annealing under such conditions does not occur. The out-diffusion of Yb-ions after annealing was a major problem for the optical application of ZnO: RE systems [18].

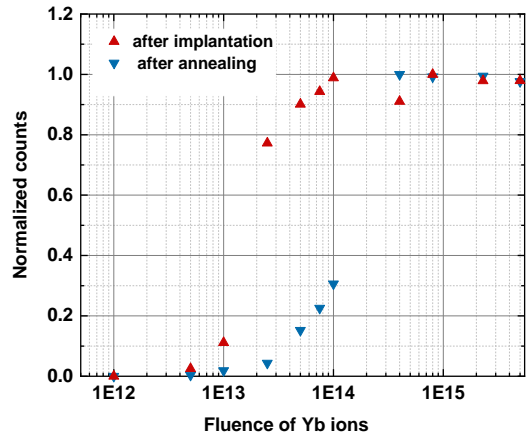

**Figure 8.** Dependence of the backscattering yield in the damage zone for β-Ga$_2$O$_3$ implanted with Yb and and post-annealed at 800 °C, in O$_2$ for 10 min.

## 4. Discussion

Adopting the consistent trend with other groups studying defects in wide bandgap semiconductors, it can be assumed that in the first stage of the defect accumulation process (region I), only point defects are created in β-Ga$_2$O$_3$ bombarded with Yb-ions. Every produced displaced atom creates stress in its vicinity, which can be released by their agglomeration. Thus, an increase of displaced atoms concentration leads to the formation of small point defect clusters which induces the growth of tensile strain in ion-bombarded ($\bar{2}$01) β-Ga$_2$O$_3$ (region II) [19,20, 27-29].

In the majority of cases, in semiconductors the relief of strain is the main driving force for amorphization [35]. Assuming that also in the β-Ga$_2$O$_3$ case, the observed rapid increase of RDA and DIS types defects in region II points at elastic deformation of the crystal lattice [28]. Above the yield point, the plastic deformation of the crystal takes place and the dislocations become mobile. Thus, the saturation of DIS-type of defects in region III is most likely associated with the formation of dislocation tangles rather than the occurrence of ideal dislocated regions.

The origin of the saturation of the RDA-type defect at the random level in region III is not fully clear. Some groups associate this, indeed, with the amorphization phase. Lorenz *et al.* [22] have observed that in ($\bar{2}$01) β-Ga$_2$O$_3$ implanted with Eu, the amorphized zone within the implanted layer begins to form for a fluence of $1 \times 10^{15}$ at/cm$^2$, which is two orders of magnitude higher than in our case. On the other hand, the recent literature data quite clearly demonstrated ion-induced phase transitions observed for the (010) oriented β-Ga$_2$O$_3$ with the high-density of radiation defects (ion fluence about $1 \times 10^{16}$ at/cm$^2$). The discussion, if it is phase transition from monoclinic (β- Ga$_2$O$_3$) to orthorhombic (κ- Ga$_2$O$_3$) or cubic (γ- Ga$_2$O$_3$) phase is ongoing [35-36]. However, it should be noticed that this crystalline-to-crystalline structural transformation results in the relief of the accumulated strain, so in a way, it is similar to the amorphization process. Moreover, according to Azarov *et al* [24], the phase transitions suppress amorphization and so they can influence each other.

Unfortunately, for β-Ga$_2$O$_3$ orientations other than (010) and smaller ion fluences, the results are not straightforward [22,37]. However, the Yb ion-induced phase transitions in ($\bar{2}$01) oriented β-Ga$_2$O$_3$ cannot be excluded, because such a crystalline-to-crystalline structural transformation is related to the coverage of the ($\bar{2}$01) channel, which also results in the increase of the backscattering yield observed in the aligned RBS/c spectra in the damage zone, similar to the displaced atoms issue blocking the channels. It should be highlighted that, apart from displaced atoms and the new phase, the line defects with edges perpendicular to the surface are also visible in the RBS/c technique as channel blocker defects, as shown in ref. [27]. Thus, the creation of basal loops may be another possible explanation for the increase in RDA in region III.

The formed RDA-types defect structure in region III would have seemed stable, if not for the result obtained for the fluence of $4 \times 10^{14}$ at/cm$^2$, where the observed RDA damage peak became bimodal and the level of RDA concentration in the region near the surface is reduced. As shown in Figure 5, for the higher used fluences, the RDA damage level reaches a saturation at 100% again and it starts to expand into the depths, while the deeper RDA-damage zone seems to disappear. The different reactions of these two damage zones to radiation could be explained only as two different RDA- types of defects. Initially, up to the fluence of $1 \times 10^{14}$ at/cm$^2$, two types of defects must develop at the same depth region, and observed saturation at 100% of RDA is the sum of their contributions. The growing concentration of one part of RDA-type of defects with the ion bombardment fluence inevitably causes the increase of stress, and this is probably the driving force behind the migration of one part of RDA-type defects (or it is pushing out) towards the depth. Above $1 \times 10^{14}$ at/cm$^2$, the transformation process is probably finished, therefore, for the fluence of $4 \times 10^{14}$ at/cm$^2$, we start to observe both RDA-type defects separately. This phenomenon requires additional extensive studies. However, as shown in our other work [38], the defect structure formed near the surface by the Yb-ions fluence higher than $4 \times 10^{14}$ at/cm$^2$, is resistant to annealing, while this deeper one disappears after annealing (see also Figure 7). It suggests that the deeper damage zone induced by ion implantation is related to the new crystalline phase of Ga$_2$O$_3$, which according to Anber *et. al* [39] is a process reversible by annealing. In turn, the damaged region closer to the surface could be associated with the amorphization [22] or other stable form of defect structure, which is visible in the RBS/c technique as a channel blocker defect.

The sudden increase in DIS concentration observed in region IV (Figure 6) indicates another evolution of dislocations tangles into stacking faults or lattice polygonization, as they are more complicated and energetically more favorable forms of extended defects. However, the changes occurring in this region can also be related to the high concentration of Yb ions. The solubility limit of Yb ions in Yb-implanted ZnO has been reported to be around 0.4 at% [29]. Above this value, the Yb$_2$O$_3$ phase has been observed [29, 40].

## 5. Conclusion

We have carried out a detailed study of the defect accumulation process in Yb-implanted $(\bar{2}01)$ oriented β-Ga$_2$O$_3$ single crystals using RBS/c technique, supported by computer simulations. Our small-step approach allowed clear observation of the structural defect transformation thresholds as a function of the growing fluence. The Monte Carlo simulations performed by the McChasy code show that the damage accumulation in Yb-implanted $(\bar{2}01)$ β-Ga$_2$O$_3$ develops in four stages. Taken as given in the literature, the driving force responsible for the defect structure transformations in $(\bar{2}01)$ oriented β-Ga$_2$O$_3$ is the accumulation of tensile strain. Importantly, our studies suggest that above the Yb-ions fluence of $1 \times 10^{13}$ ions/cm$^2$ (0.14 dpa), at least two different RDA-type defects start to develop simultaneously as a result of elastic deformation of the crystal in the damaged zone. Above the plastic deformation threshold ($1 \times 10^{14}$ Yb ions/cm$^2$ = 0.54 dpa), the two distinct damage zones become visible, indicating the migration (or pushing) of one of them into the depth. The deeper damage zone induced by Yb-ions implantation disappears after annealing, and thus it is probably related to the radiation-induced phase transformation of $Ga_2O_3$. The damage zone closer to the surface seems to be associated with amorphization, but it is contrary to the theoretical predictions about the β-$Ga_2O_3$ radiation resistance. Therefore a true understanding of the nature of the created defect structure needs further investigations. The present studies have also shown that most of the implanted Yb ions occupy the interstitial lattice site positions along the ($\bar{2}01$) direction in β-$Ga_2O_3$, and their position remains unchanged after RTA. Additionally, no diffusion of Yb-ions to the surface after annealing is observed.

**Acknowledgments**

The work was performed within the international project co-financed by the funds of the Minister of Science and Higher Education in the years 2021-2023; contract No. 5177/HZDR/2021/0 and Helmholtz-Zentrum Dresden-Rossendorf (20002208-ST). Part of the research was supported also by the NCN project UMO-022/45/B/ST5/02810.